\documentclass[multphys,vecphys]{svmult}

\usepackage{makeidx}         % allows index generation
\usepackage{graphicx}        % standard LaTeX graphics tool
\usepackage{multicol}        % used for the two-column index
\usepackage[bottom]{footmisc}% places footnotes at page bottom
\makeindex             % used for the subject index
\begin{document}

\title*{A User--oriented Comparison of the Techniques for 3D Spectroscopy}
\author{Sperello di Serego Alighieri \\
% \and
% Name of Author\inst{2}}
\institute{INAF -- Osservatorio Astrofisico di Arcetri, Largo E. Fermi 5, Firenze, Italy}
\texttt{sperello@arcetri.astro.it}}
% \and Name and Address of your Institute \texttt{name@email.address}}
\maketitle

\section{Introduction}
\label{dis:sec1}
% Always give a unique label
% and use \ref{<label>} for cross-references
% and \cite{<label>} for bibliographic references
% use \sectionmark{}
% to alter or adjust the section heading in the running head
3D spectroscopy attempts to get closer to the fundamental goal of astronomical
observing techniques, which is to {\bf record the
direction, wavelength, polarization state and arrival time for every incoming
photon over the largest field of view}. In fact
using 3D spectroscopy, the wavelength and the incoming direction in a 2D field
of view are recorded in a (x,y,$\lambda$) data cube, in contrast with standard 
techniques which either do imaging over a 2D field, or spectroscopy along a 1D slit.
There are two main ways of doing 3D spectroscopy: the best one is to
simultaneously record both direction and wavelength, while in the other one 
these are not recorded at the same time, but scanning in one of the 3 dimensions 
is required. Clearly the latter throws away some of the incoming photons, 
therefore requiring longer exposure times, and has problems with variable 
observing conditions. Nevertheless it can be useful for some particular
applications.

\section{Simultaneous Techniques}
\label{dis:sec2}

{\bf Integral field spectroscopy (IFS)} rearranges over a 2D detector the spectra coming from every pixel in a 2D
field of view. Therefore it provides a straightforward way to fill the 
(x,y,$\lambda$) data cube. The rearrangement of the spectra can be done with 
microlens arrays (e.g. SAURON \cite{dis:bac01}), fibre--lenslet arrays (e.g. 
IFS in GMOS \cite{dis:all02}), or image slicers (e.g. SINFONI
\cite{dis:eis03}). It is becoming a popular technique, since most modern
spectrographs on large telescopes have IFS capability.
IFS in general has the advantage of providing great flexibility in the choice
of the spectral resolution and wavelength range and of being easily fed by 
adaptive optics systems.  The main disadvantage is the limited field of view, 
since the number of spatial elements is limited by the number of pixels along 
one side of the detector. This drawback is at least partially
overcome by an obvious development of IFS: the field of view,
particularly for IFS using fibre--lenslet 
arrays, can be separated in several disjoint regions, for example to cover several 
galaxies in a cluster. Examples of this development are the multiple IFS of 
GIRAFFE \cite{dis:flo04}, and the even more flexible programmable IFS concept 
\cite{dis:bla04}.
Because of its flexibility, IFS is suited for a large number of applications,
from kinematical studies of the Galactic centre to stellar population and 
kinematical studies of distant galaxies \cite{dis:gil05}.

Also {\bf slitless spectroscopy} is capable of simultaneously recording a (x,y,$\lambda$)
data cube: originating from the objective prism technique, used on Schmidt 
telescopes for more than fifty years, it is easily implemented in modern imaging 
(focal reducer) spectrographs by removing the slit. Therefore it records spectra 
of all objects over the whole field of view, which can be quite large, like the 
14'x14' field of VIMOS \cite{dis:lef03}.
The disadvantages are the high sky background, since on every detector pixel
the sky is integrated over the whole wavelength range, and the overlap of spectra 
in the dispersion direction. Still this technique is particularly useful for 
surveys and searches of special objects, when the sky background is very low, like in
space or in small atmospheric windows.
For example it has been successfully used with ACS on the HST for GRAPES, a
spectroscopic survey of the Hubble Ultra Deep Field down to an AB magnitude limit 
of $z=27.2$, leading to the discovery of a large number of emission line objects over 
a huge redshift range, like AGN and Lyman $\alpha$ galaxies \cite{dis:pir04}. 
In this case the effects of the spectra overlap has
been substantially reduced by taking spectra at various position angles.
An example of the use of slitless spectroscopy from the ground is the
search for Lyman $\alpha$ emitters at z=6.5 in the atmospheric window centred at 
915 nm. In this case the spectral range can be limited to the 20 nm width of the 
window by using a narrow-band filter. Therefore both the sky emission and the 
spectra overlap are greatly reduced and very faint emission line objects can be 
found down to a line flux of $2\times 10^{-17} erg\ cm^{-2} s^{-1}$ \cite{dis:kur04}. 
% A successful application of this 
% technique is shown by the slitless spectrum in fig. \ref{fig:1}, obtained with FORS2 on the 
% VLT, which shows a Lyman $\alpha$ galaxy at z=6.518 \cite{dis:kur04}. Compared to 
% emission-line searches with narrow- and broad-band filters,
% slitless spectroscopy is equally efficient, and has the
% advantage of providing a real spectrum, from which the exact redshift and
% line profile can be derived.
% \begin{figure}
% \centering
% \includegraphics[height=5cm]{Lyaslitless.ps}
% \caption{Small section of a slitless spectrum obtained with FORS2 at the VLT,
% showing a Lyman $\alpha$ galaxy at z=6.518 \cite{dis:kur04}. Other
% emission--line objects are clearly visible.}
% \label{fig:1}       % Give a unique label
% \end{figure}

{\bf Energy--resolving detectors}
are imaging arrays where each pixel has some energy resolution. Therefore
these are true 3D devices capable of simultaneously recording the (x,y,$\lambda$) 
data cube, and no spectrograph is necessary.
Being mostly photon-counting detectors, they also have a very good temporal resolution. 
Their main disadvantages are the very limited spectral resolution and field of view.
Two different technological approaches are being explored in the optical range:
the Superconducting Tunnel Junctions (STJ \cite{dis:per93}) and the 
superconducting transition-edge sensors \cite{dis:cab98}.
Currently STJ detectors using tantalum metal films have a good quantum
efficiency in the optical range (about 70\%), but have a resolution 
$\lambda\over{\Delta\lambda}$ of only about 20 and a total number of pixels of
about 100. The latter could reasonably be increased to 10000. High speed 
energy-resolved observations of rapidly variable
stars and optical pulsars have been obtained with
STJ detectors, and their use as order sorters in an intermediate resolution
spectrograph (no 3D) has been investigated \cite{dis:cro03}.

Although purists may not consider {\bf multi--object spectroscopy} as true 3D spectroscopy, since
it does not completely cover a 2D field of view, it does however produce spectra 
of many objects in a large field, and very suitably fulfils the needs of many 
applications, making it the most popular 3D technique. Practically all telescopes have
MOS instruments, using either a fibre positioner coupled to a spectrograph, movable 
slitlets, or a multi-aperture plate. The latter implementation has advantages in terms
of better sky subtraction and throughput than fibres, and a larger number of
objects and better positioning flexibility than slitlets. A good example is VIMOS 
on the VLT \cite{dis:lef03}, which is capable of simultaneously recording spectra
of 1000 objects over a 14'x14' field of view. The disadvantages are that
it requires prior imaging (and mask preparation), that objects have to be 
preselected (not good for object searches), and that it is not capable of a complete 
2D coverage of extended objects.

\section{Scanning Techniques}
\label{dis:sec3}

{\bf Tunable imaging filters} cannot simultaneously record the data cube, but require
scanning in wavelength. The most used in astronomy is the Fabry--Perot 
filter, which uses interference between two
glass plates \cite{dis:bla00}. They
have very good imaging capability, a large field of view and good spectral
resolution. They suffer from the so--called phase
problem: the central wavelength is not constant over the
field of view. Therefore reconstructing the (x,y,$\lambda$) data cube is not 
straightforward.
Fabry--Perot filters have been used for a large number of applications mostly
on nearby galaxies and nebulae.

{\bf Imaging Fourier Transform Spectroscopy (IFTS)} is a special technique using the interference of two optical beams.
Although it requires several exposures by scanning
a movable mirror, and the reconstruction of the (x,y,$\lambda$) data cube is not
straightforward, but requires heavy computation,
nevertheless the scanning does not imply any loss of photons, which are all
recorded over the full field of view and wavelength range \cite{dis:ben00}.
A disadvantage compared to the simultaneous 3D techniques, like the IFS, is that
the readout noise affects the final data cube not just once, but a number of times 
equivalent to the number of spectral elements. Also each spectral element suffers 
the sky noise of the whole bandpass. Therefore IFTS is competitive when a reduced
number of spectral elements is required over
a large field, as, for example in kinematic studies of the Galactic centre. One
of the few examples of IFTS used in astronomy is BEAR on the CFHT \cite{dis:mai00}.

{\bf Scanning long--slit spectroscopy} does not require a new instrument, but uses a normal
long-slit spectrograph. It does not simultaneously fill the 
data cube, but can be used for very elongated objects, like edge-on 
galaxies, when only coarse information is required in the second spatial direction.

\section{Selection of the Most Suitable Technique}
\label{dis:sec4}

Although advantages and disadvantages can be found (see Table 1), it is hard to
say which technique is best. One has rather
to find the technique which most efficiently fills the (x,y,$\lambda$) data cube
for each specific application. In most cases the data cube is largely empty.
However it is exactly in these empty spaces that one can make serendipitous
discoveries.

\begin{table}
\centering
\caption{Synopsis of the techniques for 3D spectroscopy}
\label{dis:tab1}       % Give a unique label
\begin{tabular}{llll}
\hline\noalign{\smallskip}
Technique & Advantages & Disadvantages & Applications \\
\noalign{\smallskip}\hline\noalign{\smallskip}
IFS & Simoultaneous (x,y,$\lambda$) & Limited f.o.v. &
Galactic centre, \\
 & Spectral flexibility & & distant galaxies, etc. \\
 & Disjoint regions possible & & \\
\hline
Slitless spectr. & Simoultaneous (x,y,$\lambda$) & High sky background & Surveys
\\
 & Normal spectrograph & Spectra overlap & Searches for objects \\
 & Large f.o.v. & & \\
\hline
Energy--res. det. & Simoultaneous (x,y,$\lambda$) & Very limited
$\lambda\over{\Delta\lambda}$ & Rapid variables \\
 & No spectrograph & Very limited f.o.v. & Optical pulsars \\
 & Good temporal res. & Under development & \\
 & Good efficiency & & \\
\hline
MOS & Normal spectrograph & 2D field not covered & Large redshift surveys \\
 & Spectral flexibility & Prior imaging & \\
 & Very large f.o.v. & Mask preparation & \\
\hline
Tunable im. filters & Large f.o.v. & Scanning in $\lambda$ & Nearby galaxies \\
 & Good spectral res. & Variable central $\lambda$ & Nebulae \\
\hline
IFTS & No loss of photons & Scanning required & Galactic centre \\
 & & High sky background & \\
 & & Heavy computations & \\
\hline
Long slit scanning & Normal spectrograph & Scanning required & Very elongated
objects\\

\noalign{\smallskip}\hline
\end{tabular}
\end{table}
%

%
%
% BibTeX users please use
% \bibliographystyle{}
% \bibliography{}

\begin{thebibliography}{}
\bibitem[1]{dis:all02} J. Allington--Smith, G. Murray, R. Content et al: PASP
\textbf{114}, 892 (2002)
\bibitem[2]{dis:bac01} R. Bacon, Y. Copin, G. Monnet et al: MNRAS \textbf{326}, 23
(2001)
\bibitem[3]{dis:ben00} C.L. Bennett: A.S.P. Conf. Ser. Vol. 195, p. 58 (2000)
\bibitem[4]{dis:bla00} J. Bland--Hawthorn: A.S.P. Conf. Ser. Vol. 195, p. 34 (2000)
\bibitem[5]{dis:bla04} J. Bland--Hawthorn, A. McGrath, W. Saunders et al: Proc. SPIE
Vol. 5492, p. 242 (2004)
\bibitem[6]{dis:cab98} B. Cabrera, R.M. Clarke, P. Colling et al: Appl. Phys. Let.
\textbf{73(6)}, 735 (1998)
\bibitem[7]{dis:cro03} M. Cropper, M. Barlow, M.A.C. Perryman et al: MNRAS
\textbf{344}, 33 (2003)
\bibitem[8]{dis:eis03} F. Eisenhauer, R. Abuter, K. Bickert et al: Proc. SPIE Vol.
4841, p. 1548 (2003)
\bibitem[9]{dis:flo04} H. Flores, M. Puech, F. Hammer et al: A\&A \textbf{420}, L31
(2004)
\bibitem[10]{dis:gil05} S. Gillessen, R. Davies, M. Kissler--Patig et al: The Messenger
\textbf{120}, 26 (2005)
\bibitem[11]{dis:kur04} J.D. Kurk, A. Cimatti, S. di Serego Alighieri et al: A\&A
\textbf{422}, L13 (2004)
\bibitem[12]{dis:lef03} O. Le F\`evre, M. Saisse, D. Mancini et al: Proc. SPIE Vol.
4841, p. 1670 (2003)
\bibitem[13]{dis:mai00} J.P. Maillard: A.S.P. Conf. Ser. Vol. 195, p. 185 (2000)
\bibitem[14]{dis:per93} M. Perryman, C. Foden, A. Peacock et al: Nuc. Inst. Meth.
A \textbf{325}, 319 (1993)
\bibitem[15]{dis:pir04} N. Pirzkal, C. Xu, S. Malhotra et al: ApJS \textbf{154}, 501
(2004)

\end{thebibliography}
%
% Non-BibTeX users please follow the syntax
% the syntax of "referenc.tex" for your own citations
% \input{referenc}
%%%%%%%%%%%%%%%%%%%%%%%%%%%%%%%%%%%%%%%%%%%%%%%%%%%%%%%%%%%%%%%%%%%%%%  }

%%%%%%%%%%%%%%%%%%%%%%%%%%%%%%%%%%%%%%%%%%%%%%%%%%%%%%%%%%%%%%%%%%%%%%

\printindex
\end{document}